\documentclass[letterpaper, 10 pt, conference]{ieeeconf} 
\IEEEoverridecommandlockouts  
\usepackage{amsmath,amssymb,amsfonts}
\usepackage{algorithmic}
\usepackage{algorithm}
\usepackage{graphicx}
\usepackage{textcomp}
\usepackage{xcolor}
\usepackage{bm}
\usepackage{mathtools}
\usepackage{tikz}
\usepackage{cite}
\usepackage{booktabs}
\usepackage{newtxtext}

\usepackage{subcaption}
\usepackage{hyperref}
\usepackage{url}
\usepackage{xcolor}
\hypersetup{
    colorlinks=true,
    linkcolor=blue,
    citecolor=blue,
    urlcolor=blue 
}

\DeclareMathOperator*{\argmin}{arg\,min}

\newcommand{\R}{\mathbb{R}}
\newcommand{\baru}{\bar{u}}

\newtheorem{remark}{Remark}

\begin{document}

\title{\LARGE \bf DM-MPPI: Datamodel-Based Sample Pruning for Model Predictive Path Integral Control}

\author{
Jiachen Li$^{1}$,  Duan Xu$^{1}$, Shihao Li$^{1}$,  Soovadeep Bakshi$^{1}$, and Dongmei Chen$^{1}$%
\thanks{$^{1}$All authors are with the Department of Mechanical Engineering, The University of Texas at Austin, Austin, TX 78712, USA.
        {\tt\small \{jiachenli, xu.duan, shihaoli01301,  soovadeepbakshi, dmchen\}@utexas.edu}}%
}

\maketitle

\begin{abstract}
Sampling-based controllers like Model Predictive Path Integral (MPPI) generate hundreds of trajectory rollouts every control cycle, but many of these rollouts barely affect—or even hurt—the final control action. Figuring out which samples actually matter is, at its core, a data attribution problem: each rollout acts like a training example that pushes the controller's output one way or another, depending on whether it is included. We make this connection concrete by adapting the Datamodels framework, which learns how much each sample matters through regression over random subsets, to MPPI. During an offline phase, we compute influence scores across random subsets of rollouts and train a small neural network to predict influence from cost features—allowing real-time pruning without any online regression. An adaptive constraint penalty also adjusts obstacle avoidance strength based on how often sampled trajectories violate constraints. We validate the method on a kinematic bicycle model driving through two obstacle-filled tracks.
\end{abstract}

\section{Introduction}

Model Predictive Path Integral (MPPI) control has become a popular sampling-based approach for nonlinear systems~\cite{williams2016aggressive,williams2017model}, achieving real-time performance by rolling out many trajectories in parallel and blending them through importance-weighted averaging~\cite{williams2018information}. However, two key challenges remain: a large fraction of the sampled trajectories has little effect on the final control action, leading to wasted computation, and soft constraint penalties~\cite{zeilinger2014soft} make it unclear how constraint-violating samples shape the output. A natural question is whether we can identify which samples actually matter and focus the computation on those. We argue that this is naturally a \emph{data attribution} problem — each rollout is like a training example that, when included or left out, shifts the control update in a different direction. The Datamodels framework~\cite{ilyas2022datamodels} offers a clean solution by learning per-example influence coefficients through subset regression. We bring this idea into MPPI by moving the expensive regression to an offline stage and training a small neural network to predict influence at deployment time.

Path integral methods recover optimal control laws by exponentially weighting sampled trajectories~\cite{theodorou2010generalized}. Williams et al.~\cite{williams2016aggressive,williams2017model} introduced MPPI for real-time GPU-parallel control and showed its effectiveness in aggressive autonomous driving~\cite{williams2018information}. Follow-up work has since improved robustness~\cite{gandhi2021robust}, smoothness~\cite{kim2022smooth}, and sample efficiency~\cite{mohamed2025umppi}. On the constraint side, researchers have explored CVaR-based risk awareness~\cite{yin2022ramppi}, Control Barrier Functions~\cite{yin2023shieldmppi}, tube-based formulations~\cite{balci2022ccsmppi}, and safe importance sampling~\cite{gandhi2024safe}. Deep learning intersects with control through data-driven synthesis~\cite{depersis2020formulas,lusch2018deep,wu2019machine} and neural certificates~\cite{dawson2023safe}, with interpretability drawing growing attention~\cite{wang2023measuring}. In the machine learning literature, data attribution has been studied through influence functions~\cite{koh2017understanding}, Data Shapley~\cite{ghorbani2019data}, TracIn~\cite{pruthi2020estimating}, the Datamodels framework~\cite{ilyas2022datamodels}, and TRAK~\cite{park2023trak}. To our knowledge, no prior work has applied data attribution to sampling-based control.

Our contributions are threefold. First, we recast MPPI sample selection as a data attribution problem and adapt the Datamodels framework through an offline--online setup: influence coefficients are fitted via subset regression during training and predicted by a neural network at runtime. Second, we compare nine methods on a kinematic bicycle model across two track geometries with obstacles. Keeping only $20\%$ of samples through informed pruning substantially lowers tracking error compared to unpruned MPPI, while random pruning is worse than even the smallest sample budget. Adding more samples shows diminishing returns, but pruning down to a few well-chosen samples consistently beats the full unpruned budget. Third, we show that the learned influence predictor with adaptive mechanisms (DM-adapted) achieves the fewest boundary violations across both tracks, significantly outperforming heuristic pruning methods that tend to keep trajectories running tight along obstacles—whereas influence-aware selection naturally avoids putting too much weight on samples near the constraint boundary.

The rest of this paper is organized as follows. Section~II covers background on MPPI and the Datamodels framework. Section~III presents the proposed framework. Section~IV evaluates against baselines on two track geometries. Section~V concludes.
\section{Preliminaries}

\subsection{Model Predictive Path Integral Control}

Consider a discrete-time nonlinear system with a control-affine structure.
\begin{equation}
    x_{t+1} = F(x_t, u_t + \epsilon_t)
\end{equation}
where $x_t \in \R^n$ is the state, $u_t \in \R^m$ is the control input, $F: \R^n \times \R^m \to \R^n$ is the dynamics map, and $\epsilon_t \sim \mathcal{N}(0, \Sigma)$ represents zero-mean Gaussian control noise with covariance $\Sigma \in \R^{m \times m}$.

The total trajectory cost consists of multiple components
\begin{equation}
    S(\tau) = S_{\text{goal}}(\tau) + S_{\text{control}}(\tau) + \rho \cdot S_{\text{violation}}(\tau)
    \label{eq:total_cost}
\end{equation}
where the goal cost captures task objectives $S_{\text{goal}}(\tau) = \phi(x_T) + \sum_{t=0}^{T-1} q(x_t)$, the control cost follows the standard MPPI formulation $S_{\text{control}}(\tau_k) = \lambda \sum_{t=0}^{T-1} \baru_t^\top R \, \epsilon_{t,k}$ (a cross-term between the nominal control and the perturbation~\cite{williams2017model}), and the violation cost captures soft constraint penalties $S_{\text{violation}}(\tau) = \sum_{t=0}^{T-1} c(x_t)$ with $\rho > 0$ controlling constraint importance.

Given $K$ sampled trajectories $\{\tau_k\}_{k=1}^K$, each generated by applying perturbed controls $u_t^k = \baru_t + \epsilon_t^k$, the optimal control update is
\begin{equation}
    u_t^* = \baru_t + \sum_{k=1}^K w_k \epsilon_t^k
    \label{eq:mppi_update}
\end{equation}
where the importance weights follow the Gibbs distribution
\begin{equation}
    w_k = \frac{\exp(-S(\tau_k)/\lambda)}{\sum_{j=1}^K \exp(-S(\tau_j)/\lambda)}.
    \label{eq:mppi_weights}
\end{equation}
The temperature $\lambda > 0$ governs weight concentration \cite{williams2017model}.

\subsection{The Datamodels Framework}

The Datamodels framework \cite{ilyas2022datamodels} provides a principled approach to understanding how individual data points influence model outputs. The core idea is to learn a simple predictor that maps training set composition to model behavior, enabling post-hoc attribution of predictions to specific training examples.

For a dataset $\mathcal{S} = \{z_1, \ldots, z_n\}$ and a model output of interest $f(x; S')$ trained on subset $S' \subseteq \mathcal{S}$, a linear datamodel takes the form
\begin{equation}
    g_\theta(\mathbf{1}_{S'}) = \theta^\top \mathbf{1}_{S'} + \theta_0
\end{equation}
where $\mathbf{1}_{S'} \in \{0,1\}^n$ is a binary vector indicating which data points are included in the subset $S'$. The coefficient $\theta_j$ quantifies the marginal influence of data point $z_j$ on the output: a positive $\theta_j$ indicates that including $z_j$ increases the predicted output, while a negative $\theta_j$ indicates a decreasing effect.

Training proceeds by constructing many random subsets $\{S'_i\}_{i=1}^M$, each formed by including each data point independently with a probability $\alpha \in (0,1)$. For each subset, the true model output $f(x; S'_i)$ is computed. The datamodel parameters are then learned via LASSO regression \cite{park2023trak}:
\begin{equation}
    \theta^* = \argmin_\theta \frac{1}{M}\sum_{i=1}^M \left(g_\theta(\mathbf{1}_{S'_i}) - f(x; S'_i)\right)^2 + \mu\|\theta\|_1
\end{equation}
The $\ell_1$ penalty encourages sparsity, reflecting the intuition that only a subset of data points meaningfully influence any given prediction.

Unlike influence functions \cite{koh2017understanding}, which rely on local approximations around the full training set, datamodels directly learn global input-output relationships across the space of possible training subsets.

\subsection{Motivation for Linear Influence Models in MPPI}
The MPPI control update~\eqref{eq:mppi_update} is a weighted sum of perturbations, where inclusion or exclusion of a sample changes both the numerator and the normalizing denominator of the weights. The control update is therefore not linear in the inclusion vector $\mathbf{b} \in \{0,1\}^K$, and there is no formal guarantee that the LASSO regression will recover accurate influence coefficients. Nonetheless, empirically we find that the linear datamodel achieves $R^2 \approx 0.6$ when predicting the first-step control update from sample inclusion (Section~IV). We treat this as a practical heuristic: the moderate $R^2$ indicates that a linear model captures some structure in sample influence, and the MLP predictor trained on these coefficients inherits both the signal and the approximation error. Whether the resulting pruning criterion is better than simple cost-based heuristics is an empirical question that we investigate in Section~IV.

\section{Method: MPPI-Datamodel Framework}

\begin{figure}[t]
\centering
\includegraphics[width=0.48\textwidth]{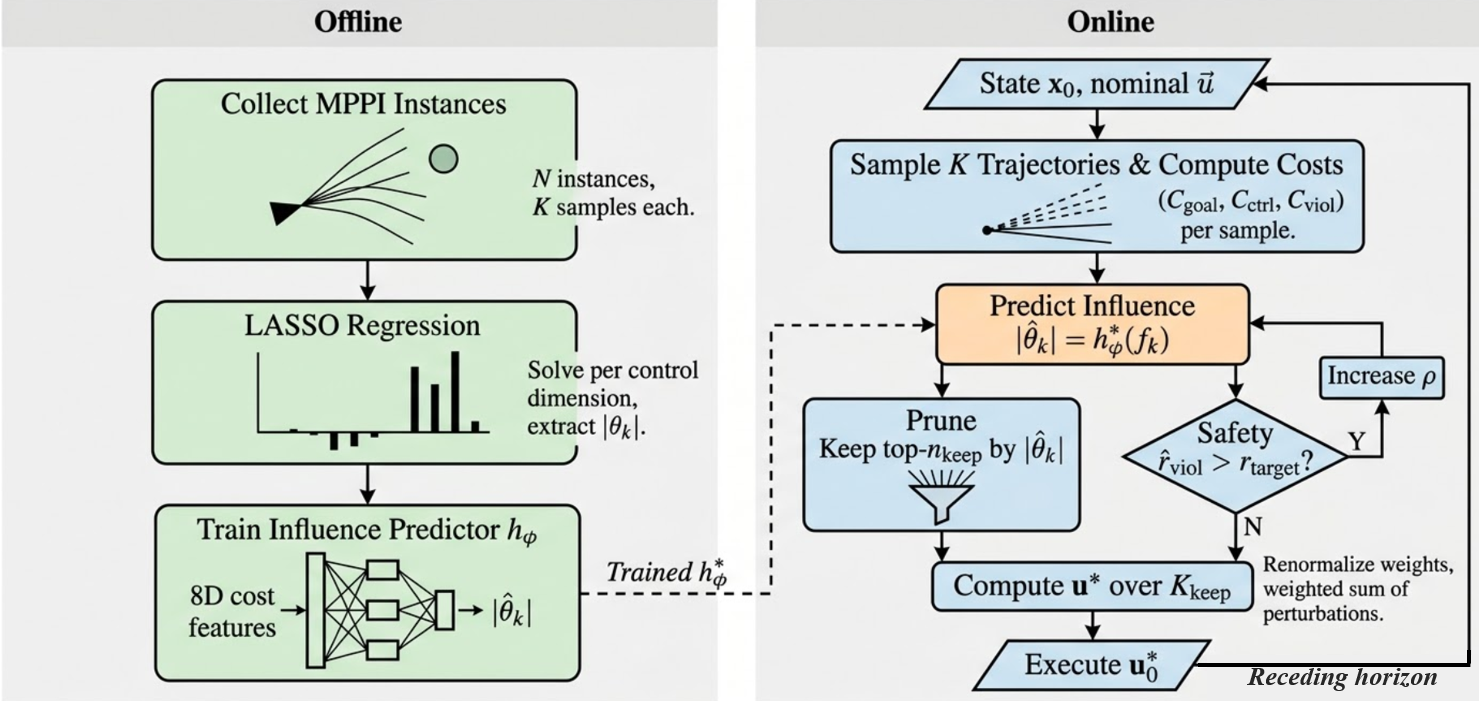}
\caption{DM-MPPI framework.}
\label{fig:workflow}
\end{figure}

\subsection{Problem Formulation}

Consider an MPPI instance with $K$ sampled trajectories $\mathcal{T} = \{\tau_1, \ldots, \tau_K\}$. Each trajectory $\tau_k$ incurs a total cost $C_k$ composed of goal cost, control cost, and constraint violation cost:
\begin{equation}
    C_k = C_{\text{goal},k} + C_{\text{ctrl},k} + \rho \cdot C_{\text{viol},k}
\end{equation}
where $\rho > 0$ is the soft constraint penalty coefficient. For a subset $\mathcal{T}' \subseteq \mathcal{T}$, the subset-restricted control update is computed by renormalizing importance weights over $\mathcal{T}'$:
\begin{equation}
    w_k(\mathcal{T}') = \frac{\exp(-C_k/\lambda)}{\sum_{j: \tau_j \in \mathcal{T}'} \exp(-C_j/\lambda)}
\end{equation}
\begin{equation}
    \delta u_0(\mathcal{T}') = \sum_{k: \tau_k \in \mathcal{T}'} w_k(\mathcal{T}') \, \epsilon_{0,k}
\end{equation}
where $\epsilon_{0,k}$ is the first-step control perturbation of trajectory $\tau_k$.

To capture the relationship between sample inclusion and the control update, we define the binary inclusion vector $\mathbf{b} \in \{0,1\}^K$ where $b_k = 1$ if trajectory $\tau_k$ is included. The MPPI-Datamodel predicts the control update from sample membership via the linear model
\begin{equation}
    g_\theta(\mathbf{b}) = \theta^\top \mathbf{b} + \theta_0 \approx \delta u_0(\mathcal{T}')
    \label{eq:mppi_datamodel}
\end{equation}
where we fit separate regressions for each control dimension and average the absolute coefficients: $|\theta_k| = \frac{1}{m}\sum_{d=1}^m |\theta_k^{(d)}|$. The resulting $|\theta_k|$ quantifies the marginal influence of sample $\tau_k$ on the immediate control action. Samples with large $|\theta_k|$ disproportionately affect the first control applied to the system.

\subsection{Offline Training Phase}

We adopt an offline training paradigm to avoid repeated regression during deployment. The key insight from the datamodeling framework~\cite{ilyas2022datamodels} is that by training on many random subsets, we can recover individual sample influences even though we never isolate single samples. The offline procedure consists of three steps.

\textbf{Step 1: Data Collection.}
We collect $N$ MPPI instances across representative states sampled from the expected operating distribution. For each instance $i$, we draw $K$ trajectories using the nominal sampling distribution and record the cost components $(C_{\text{goal},k}^{(i)}, C_{\text{ctrl},k}^{(i)}, C_{\text{viol},k}^{(i)})$ for each trajectory $k$. Given a fixed penalty $\rho_0$, we compute total costs $C_k^{(i)} = C_{\text{goal},k}^{(i)} + C_{\text{ctrl},k}^{(i)} + \rho_0 \cdot C_{\text{viol},k}^{(i)}$. To ensure the learned predictor generalizes, instances should cover diverse regions of the state space, including states near constraints.

\textbf{Step 2: Datamodel Fitting.}
For each instance $i$, we learn the influence coefficients by regression on random subsets. We generate $M$ random subsets by including each trajectory independently with probability $\alpha \in (0,1)$. For each subset $j$ with an inclusion mask $\mathbf{b}_j \in \{0,1\}^K$, we compute the subset-restricted control update $\delta u_{0,j}^{(i)} \in \R^m$ by renormalizing the MPPI weights over the included trajectories and taking the weighted sum of first-step perturbations. The influence coefficients for each control dimension $d$ are obtained by solving
\begin{equation}
    \theta^{(i,d)*} = \argmin_{\theta, \theta_0} \frac{1}{M}\sum_{j=1}^M \left(\theta^\top \mathbf{b}_j + \theta_0 - [\delta u_{0,j}^{(i)}]_d\right)^2 + \mu\|\theta\|_1
    \label{eq:lasso}
\end{equation}
where $\theta_0$ is an intercept term excluded from regularization, and the $\ell_1$ penalty encourages sparsity. We then compute a single influence score per sample by averaging absolute coefficients across dimensions: $|\theta_k^{(i)*}| = \frac{1}{m}\sum_{d=1}^m |\theta_k^{(i,d)*}|$.

The key to recovering individual influences is the randomness of the subsets. For any two trajectories $k \neq k'$, their inclusion events are independent across subsets:
\begin{equation}
    \mathbb{E}[b_{j,k} \cdot b_{j,k'}] = \mathbb{E}[b_{j,k}] \cdot \mathbb{E}[b_{j,k'}] = \alpha^2
\end{equation}
This independence decorrelates the contributions of different samples, helping the regression disentangle individual effects from aggregate subset effects. The LASSO solution $\theta_k^{(i)*}$ provides an approximate influence score for each sample $k$; the quality of this approximation depends on $M$ and on the degree of nonlinearity in the control update mapping (Section~II-C).

\textbf{Step 3: Influence Predictor Training.}
We train a predictor $h_\phi$ that generalizes influence estimation across instances. A critical observation is that the influence coefficient $\theta_k^{(i)*}$ depends not only on sample $k$'s own cost, but also on the cost distribution of other samples in instance $i$. Intuitively, a trajectory with cost $C_k = 10$ has high influence if all other trajectories have costs above $100$, but low influence if other trajectories have similar costs around $10$.

To capture this context dependence, we construct an 8-dimensional feature vector:
\begin{equation}
    \hat{\theta}_k = h_\phi\left(\mathbf{f}_k^{(i)}\right)
    \label{eq:predictor}
\end{equation}
where $\mathbf{f}_k^{(i)} = [C_k,\; C_{\text{viol},k},\; z_k,\; \tilde{C}_k,\; s_k,\; \mathbf{1}_{C_{\text{viol},k}>0},\; \bar{C}^{(i)},\; \sigma_C^{(i)}]^\top$. Here $z_k = (C_k - \bar{C})/\sigma_C$ is the $z$-score, $\tilde{C}_k = (C_k - C_{\min})/\sigma_C$ is the normalized distance from the best sample, $s_k = \exp(-(C_k - C_{\min})/\sigma_C)$ is a soft exponential rank, $\bar{C}^{(i)} = \frac{1}{K}\sum_k C_k^{(i)}$ is the mean cost, and $\sigma_C^{(i)} = \text{std}(\{C_k^{(i)}\})$ is the cost standard deviation. The per-sample features ($z_k, \tilde{C}_k, s_k$) encode where sample $k$ sits in the cost distribution, while the instance statistics ($\bar{C}, \sigma_C$) provide context.

The predictor parameters are learned by minimizing the prediction error across all collected instances:
\begin{equation}
    \phi^* = \argmin_\phi \sum_{i=1}^{N} \sum_{k=1}^{K} \left( h_\phi\left(\mathbf{f}_k^{(i)}\right) - |\theta_k^{(i)*}| \right)^2
    \label{eq:predictor_loss}
\end{equation}
In practice, $h_\phi$ can be a small neural network (e.g., two hidden layers with 64 units) or even a linear model if the relationship is sufficiently simple. Algorithm~\ref{alg:offline} summarizes the complete offline training procedure.

\begin{algorithm}[t]
\caption{Offline: Influence Predictor Training}
\label{alg:offline}
\begin{algorithmic}[1]
\REQUIRE Number of instances $N$, samples per instance $K$, subsets per instance $M$, inclusion probability $\alpha$, regularization $\mu$, initial penalty $\rho_0$
\ENSURE Trained predictor $h_{\phi^*}$
\STATE Initialize dataset $\mathcal{D} \gets \emptyset$
\FOR{$i = 1$ to $N$}
    \STATE Sample state $x_0^{(i)}$ from operating distribution
    \STATE Collect $K$ trajectories, compute cost components $\{C_{\text{goal},k}^{(i)}, C_{\text{ctrl},k}^{(i)}, C_{\text{viol},k}^{(i)}\}_{k=1}^K$
    \STATE Compute total costs: $C_k^{(i)} \gets C_{\text{goal},k}^{(i)} + C_{\text{ctrl},k}^{(i)} + \rho_0 \cdot C_{\text{viol},k}^{(i)}$
    \STATE Compute importance weights: $w_k^{(i)} \gets \exp(-C_k^{(i)}/\lambda)$
    \STATE Compute instance statistics: $\bar{C}^{(i)} \gets \frac{1}{K}\sum_k C_k^{(i)}$, \quad $\sigma_C^{(i)} \gets \text{std}(\{C_k^{(i)}\})$
    \FOR{$j = 1$ to $M$}
        \STATE Sample inclusion mask: $b_{j,k} \sim \text{Bernoulli}(\alpha)$ for all $k$
        \STATE Compute subset control update: $\delta u_{0,j}^{(i)} \gets \sum_{k: b_{j,k}=1} \tilde{w}_k^{(i,j)} \, \epsilon_{0,k}^{(i)}$
        \STATE \quad with $\tilde{w}_k^{(i,j)} = \exp(-C_k^{(i)}/\lambda) / \sum_{l: b_{j,l}=1} \exp(-C_l^{(i)}/\lambda)$
    \ENDFOR
    \STATE Solve LASSO (\ref{eq:lasso}) per control dimension, average $|\theta_k^{(i)*}| = \frac{1}{m}\sum_d |\theta_k^{(i,d)*}|$
    \FOR{$k = 1$ to $K$}
        \STATE Build feature vector $\mathbf{f}_k^{(i)}$ per~\eqref{eq:predictor}
        \STATE $\mathcal{D} \gets \mathcal{D} \cup \{(\mathbf{f}_k^{(i)}, |\theta_k^{(i)*}|)\}$
    \ENDFOR
\ENDFOR
\STATE Train $h_{\phi^*}$ on $\mathcal{D}$ by minimizing (\ref{eq:predictor_loss})
\RETURN $h_{\phi^*}$
\end{algorithmic}
\end{algorithm}

\subsection{Online Inference Phase}

At runtime, the learned predictor enables efficient influence estimation without solving any regression problems. The predictor uses the decomposed costs $(C_k, C_{\text{viol},k})$ along with instance statistics as input features. Importantly, the actual MPPI control computation still operates on the total cost with the current penalty $\rho$, which can be adjusted online without retraining the predictor.

Given a new set of $K$ sampled trajectories, we compute cost components and build the 8-dimensional feature vector $\mathbf{f}_k$ for each sample as in~\eqref{eq:predictor}. The predicted influence magnitudes are
\begin{equation}
    \hat{\theta}_k = h_{\phi^*}(\mathbf{f}_k)
    \label{eq:online_predict}
\end{equation}
These predictions identify which samples contribute meaningfully to the control output.

\textbf{Sample Pruning for Efficiency.}
We retain the top $n_{\text{keep}}$ samples ranked by predicted influence magnitude:
\begin{equation}
    \mathcal{K}_{\text{keep}} = \text{top-}n_{\text{keep}}\{k : |\hat{\theta}_k|\}
\end{equation}
This fixed-size pruning step preserves the samples that matter most for the control update. The MPPI control update is then computed using only the retained samples:
\begin{equation}
    u^* = \sum_{k \in \mathcal{K}_{\text{keep}}} \tilde{w}_k \, u_k, \qquad \tilde{w}_k = \frac{\exp(-C_k/\lambda)}{\sum_{j \in \mathcal{K}_{\text{keep}}} \exp(-C_j/\lambda)}
\end{equation}
where the importance weights are renormalized over $\mathcal{K}_{\text{keep}}$.

\textbf{Adaptive Constraint Penalty for Safety.}
To adapt the constraint penalty online, we monitor the fraction of sampled trajectories that violate constraints. We define the violation rate as
\begin{equation}
    r_{\text{viol}} = \frac{1}{K}\sum_{k=1}^K \mathbb{1}[C_{\text{viol},k} > 0]
\end{equation}
and maintain an exponential moving average $\hat{r}_{\text{viol},t} = (1-\alpha_{\text{ema}}) \hat{r}_{\text{viol},t-1} + \alpha_{\text{ema}} \, r_{\text{viol},t}$ with $\alpha_{\text{ema}} = 0.1$. When $\hat{r}_{\text{viol}}$ exceeds the target, many trajectories violate constraints, suggesting the penalty $\rho$ is insufficient. We update
\begin{equation}
    \rho_{t+1} = \text{clip}\left(\rho_t + \eta \cdot (\hat{r}_{\text{viol},t} - r_{\text{target}}),\; \rho_{\min},\; \rho_{\max}\right)
    \label{eq:adaptive_rho}
\end{equation}
where $\eta > 0$ is a step size and $r_{\text{target}} \in (0,1)$ is the desired maximum violation rate. Algorithm~\ref{alg:online} presents the complete online procedure.

\begin{algorithm}[t]
\caption{Online: Datamodel-Based MPPI}
\label{alg:online}
\begin{algorithmic}[1]
\REQUIRE Trained predictor $h_{\phi^*}$, initial state $x_0$, nominal control $\bar{U}$, initial penalty $\rho$, number to keep $n_{\text{keep}}$, target violation rate $r_{\text{target}}$, step size $\eta$, EMA parameter $\alpha_{\text{ema}}$, temperature $\lambda$
\STATE $\hat{r}_{\text{viol}} \gets 0$
\FOR{each control iteration}
    \STATE Sample $K$ perturbations $\{\epsilon_k\}$, roll out trajectories from current state
    \STATE Compute cost components: $(C_{\text{goal},k}, C_{\text{ctrl},k}, C_{\text{viol},k})$ for all $k$
    \STATE Compute total costs: $C_k \gets C_{\text{goal},k} + C_{\text{ctrl},k} + \rho \cdot C_{\text{viol},k}$
    \STATE Build features $\mathbf{f}_k$ and predict influence: $\hat{\theta}_k \gets h_{\phi^*}(\mathbf{f}_k)$
    \STATE Select top $n_{\text{keep}}$ samples by $|\hat{\theta}_k|$: $\mathcal{K}_{\text{keep}}$
    \STATE Compute violation rate: $r_{\text{viol}} \gets \frac{1}{K}\sum_k \mathbf{1}[C_{\text{viol},k} > 0]$
    \STATE Update EMA: $\hat{r}_{\text{viol}} \gets (1-\alpha_{\text{ema}})\hat{r}_{\text{viol}} + \alpha_{\text{ema}} \, r_{\text{viol}}$
    \STATE Update penalty: $\rho \gets \text{clip}(\rho + \eta(\hat{r}_{\text{viol}} - r_{\text{target}}), \rho_{\min}, \rho_{\max})$
    \STATE Recompute weights over $\mathcal{K}_{\text{keep}}$: $\tilde{w}_k \gets \frac{\exp(-C_k/\lambda)}{\sum_{j \in \mathcal{K}_{\text{keep}}} \exp(-C_j/\lambda)}$
    \STATE Compute control: $u^* \gets \bar{u} + \sum_{k \in \mathcal{K}_{\text{keep}}} \tilde{w}_k \, \epsilon_k$
    \STATE Execute first control $u_0^*$, shift nominal sequence $\bar{U}$
\ENDFOR
\end{algorithmic}
\end{algorithm}

\begin{remark}[Pruning and Penalty Adaptation]
The pruning step and the penalty adaptation work independently of each other. Pruning keeps the samples with the highest predicted influence on the control update, so the weighted average is computed over fewer but more informative samples. The adaptive penalty tracks how often sampled trajectories violate constraints and adjusts $\rho$ to compensate. It is worth noting that pruning by influence magnitude does not automatically favor safe trajectories — a sample with high influence can still violate constraints (see Section~IV).
\end{remark}

\begin{remark}[Computational Complexity]
The offline phase involves $O(N \cdot M \cdot K)$ subset evaluations and $N$ LASSO solves, but it only needs to be done once. At runtime, the online phase only requires $O(K)$ forward passes through $h_{\phi^*}$ per control step, which is small compared to the cost of sampling trajectories and evaluating costs. Pruning also reduces the weighted average computation from $O(K)$ down to $O(|\mathcal{K}_{\text{keep}}|)$.
\end{remark}
\section{Experiments}

We evaluate the proposed framework on two closed-loop tracks: a heart-shaped course and a figure-8 course, with static obstacle avoidance using a kinematic bicycle model. All methods share identical dynamics and track geometry; they differ only in sample count and cost/sampling strategy. All experiments use 8 random seeds and report the mean $\pm$ standard deviation.

\subsection{Experimental Setup}

\textbf{Dynamics.} We use a kinematic bicycle model with state $x = [p_x, p_y, \psi, v]^\top$ and control $u = [a, \delta]^\top$ (acceleration, steering angle):
\begin{align}
    p_{x,t+1} &= p_{x,t} + v_t \cos(\psi_t)\,\Delta t \nonumber \\
    p_{y,t+1} &= p_{y,t} + v_t \sin(\psi_t)\,\Delta t \nonumber \\
    \psi_{t+1} &= \psi_t + \frac{v_t}{L}\tan(\delta_t)\,\Delta t \nonumber \\
    v_{t+1} &= \text{clip}(v_t + a_t\,\Delta t,\; v_{\min},\; v_{\max})
\end{align}
with wheelbase $L = 0.3$\,m, $\Delta t = 0.1$\,s, $v_{\max} = 2.5$\,m/s, $v_{\min} = 0.3$\,m/s, and maximum steering angle $|\delta| \leq 0.55$\,rad.

\textbf{Track environments.} We consider two closed-loop reference paths:

\emph{Heart track.} A modified parametric heart curve:
\begin{align}
    h_x(t) &= s \cdot 16\sin^3(t), \nonumber \\
    h_y(t) &= s \cdot (13\cos t - 3\cos 2t - 0.8\cos 3t - 0.2\cos 4t)
\end{align}
with scale $s = 0.6$, centered at $(5, 5)$\,m. The track has an arc length of $53.3$\,m, a constant width of $0.7$\,m, and ten rectangular obstacles ($0.40 \times 0.28$\,m) placed at evenly spaced arc length fractions with alternating lateral offsets.

\emph{Lemniscate track.} A lemniscate of Bernoulli:
\begin{align}
    \ell_x(t) &= \frac{a\cos t}{1 + \sin^2 t}, \qquad
    \ell_y(t) = \frac{a\sin t \cos t}{1 + \sin^2 t}
\end{align}
with $a = 4.5$, centered at $(5, 5)$\,m. The figure-8 has a self-intersection where the track narrows to $30\%$ of the base width ($0.7$\,m). Eight obstacles are placed along the path with tighter lateral offsets near the crossing.

\subsection{Baselines}
\begin{figure}[t]
    \centering
    \begin{minipage}[t]{0.48\columnwidth}
        \centering
        \includegraphics[width=\linewidth]{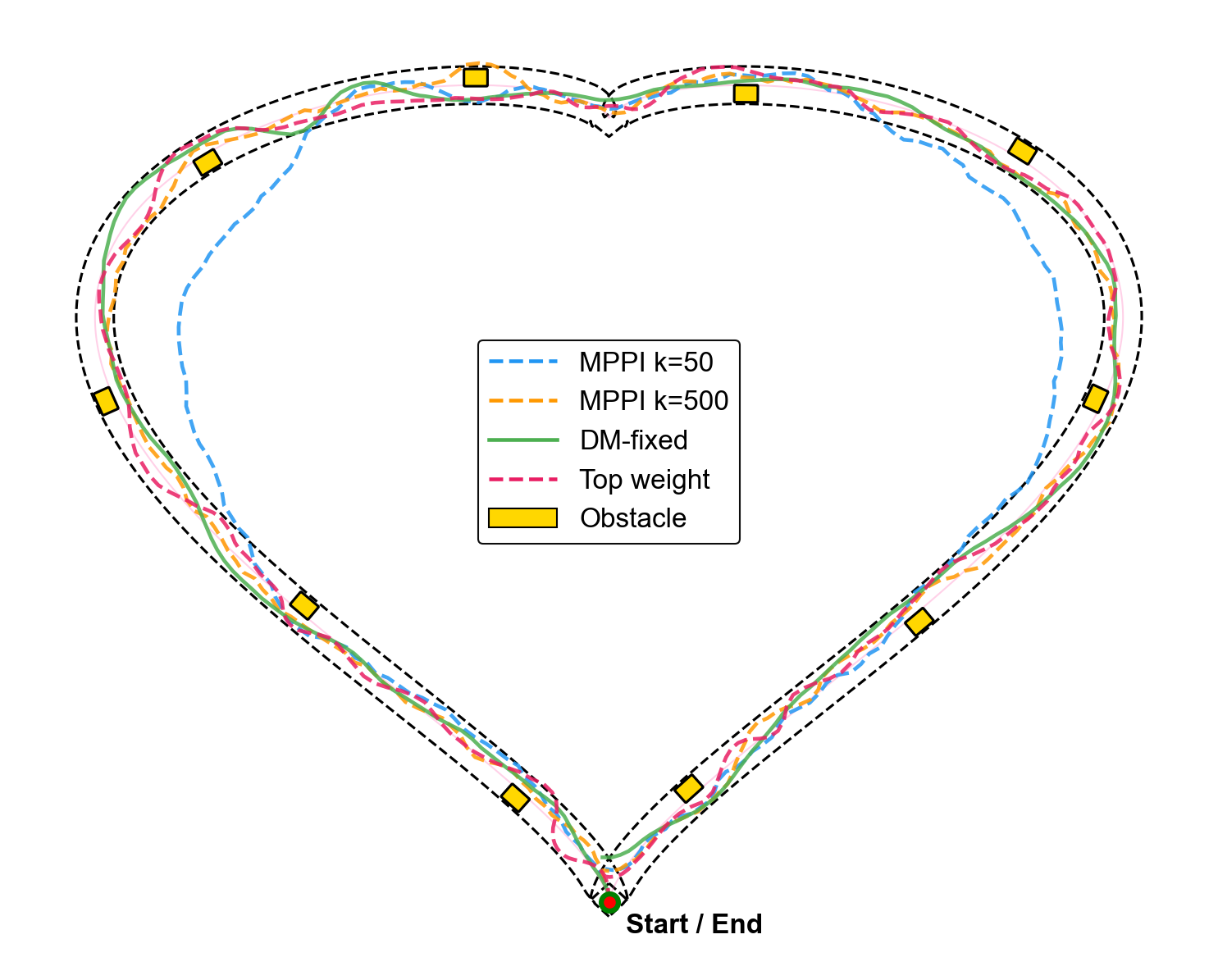}
    \end{minipage}
    \hfill
    \begin{minipage}[t]{0.48\columnwidth}
        \centering
        \includegraphics[width=\linewidth]{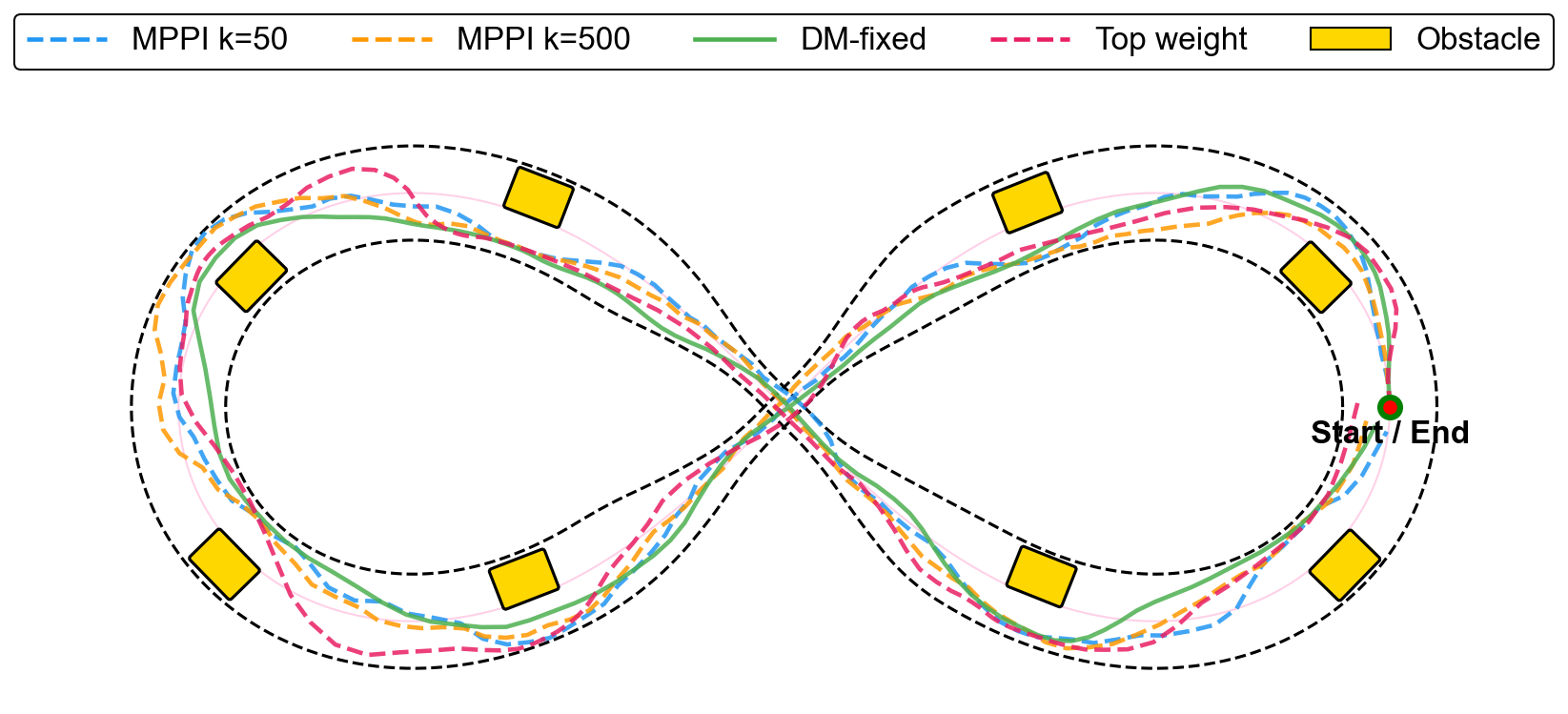}
    \end{minipage}
    \caption{Representative trajectories on the heart track (left) and lemniscate track (right).}
    \label{fig:comparison}
\end{figure}

We compare nine methods spanning standard MPPI at multiple sample budgets, heuristic pruning, and the proposed datamodel-based variants (Fig.~\ref{fig:comparison}). MPPI $K\!=\!50/100/200/500$ runs standard MPPI with i.i.d.\ samples; $K\!=\!50$ uses only quadratic centerline cost ($w_{\text{track}} = 8.0$), obstacle barriers, and a progress reward, while $K \geq 100$ adds boundary violation and heading alignment penalties. All four pruning baselines start with $K\!=\!500$ samples and keep 100. Random Pruning simply picks 100 at random, giving a lower bound on what pruning can do. Top-Weight Pruning keeps the 100 samples that already carry the most MPPI importance weight $w_k$—in other words, the trajectories that would dominate the weighted average anyway. Cost-Threshold Pruning throws out the 400 most expensive trajectories and recomputes MPPI weights over the remaining 100. Elite-CEM also keeps the 100 cheapest samples, but instead of applying MPPI weights, it just averages their perturbations uniformly, following the Cross-Entropy Method idea of working with an elite set. DM-fixed builds on MPPI $K\!=\!500$ with temporally correlated AR(1) noise ($\beta = 0.6$), 3 MPPI iterations per timestep, control smoothness penalty, curvature-aware target speed, and influence-based pruning. DM-adapted extends DM-fixed with adaptive temperature scaling, log-barrier boundary cost, look-ahead curvature-aware speed, ESS-driven noise adaptation, and elite-sample refinement. 

\subsection{Results}

\begin{table}[t]
\centering
\caption{Results across both tracks (8 seeds). Pruning methods draw $K\!=\!500$ and retain 100 samples.}
\label{tab:combined_results}
\footnotesize
\setlength{\tabcolsep}{3pt}
\begin{tabular}{@{}lccc@{}}
\toprule
\textbf{Method} & \textbf{Tracking RMSE} & \textbf{Heading RMSE} & \textbf{Boundary} \\
 & (m) & (rad) & \textbf{Viol.} \\
\midrule
\multicolumn{4}{@{}l}{\textit{Heart track (10 obstacles, arc length 53.3\,m)}} \\[2pt]
MPPI $K\!=\!50$       & $0.65 \pm 0.10$ & $0.39 \pm 0.12$ & $88.7 \pm 12.1$ \\
MPPI $K\!=\!100$      & $0.38 \pm 0.06$ & $0.33 \pm 0.08$ & $31.3 \pm 8.5$  \\
MPPI $K\!=\!200$      & $0.25 \pm 0.04$ & $0.30 \pm 0.07$ & $6.0 \pm 5.2$   \\
MPPI $K\!=\!500$      & $0.18 \pm 0.02$ & $0.28 \pm 0.01$ & $0.7 \pm 0.6$   \\
Random Prune          & $0.71 \pm 0.06$ & $0.42 \pm 0.08$ & $97.3 \pm 14.8$ \\
Top-Weight            & $0.21 \pm 0.03$ & $0.26 \pm 0.10$ & $2.0 \pm 1.7$   \\
Cost-Threshold        & $0.24 \pm 0.04$ & $0.26 \pm 0.08$ & $6.0 \pm 2.7$   \\
Elite-CEM             & $0.28 \pm 0.06$ & $0.26 \pm 0.05$ & $8.7 \pm 1.5$   \\
DM-fixed              & $0.17 \pm 0.02$ & $\mathbf{0.23 \pm 0.03}$ & ${1.2 \pm 0.4}$ \\
\textbf{DM-adapted}   & $\mathbf{0.17 \pm 0.01}$ & $0.24 \pm 0.05$ & $\mathbf{0.8 \pm 0.3}$ \\
\midrule
\multicolumn{4}{@{}l}{\textit{Lemniscate track (8 obstacles, self-intersection)}} \\[2pt]
MPPI $K\!=\!50$       & $0.58 \pm 0.10$ & $0.36 \pm 0.03$ & $62.3 \pm 10.4$ \\
MPPI $K\!=\!100$      & $0.34 \pm 0.03$ & $0.30 \pm 0.02$ & $22.7 \pm 6.8$  \\
MPPI $K\!=\!200$      & $0.22 \pm 0.02$ & $0.27 \pm 0.02$ & $5.3 \pm 4.1$   \\
MPPI $K\!=\!500$      & $0.16 \pm 0.01$ & $0.25 \pm 0.01$ & $1.3 \pm 0.5$   \\
Random Prune          & $0.64 \pm 0.07$ & $0.39 \pm 0.14$ & $71.0 \pm 11.6$ \\
Top-Weight            & $0.25 \pm 0.07$ & $0.23 \pm 0.07$ & $8.3 \pm 1.2$   \\
Cost-Threshold        & $0.31 \pm 0.05$ & $0.23 \pm 0.12$ & $9.4 \pm 4.2$   \\
Elite-CEM             & $0.39 \pm 0.08$ & $0.24 \pm 0.05$ & $7.0 \pm 3.0$   \\
DM-fixed              & $0.19 \pm 0.03$ & $\mathbf{0.21 \pm 0.08}$ & ${2.1 \pm 0.7}$ \\
\textbf{DM-adapted}   & $\mathbf{0.16 \pm 0.02}$ & $0.22 \pm 0.05$ & $\mathbf{1.0 \pm 0.5}$ \\
\bottomrule
\end{tabular}
\end{table}

Table~\ref{tab:combined_results} presents results across both tracks. The clearest takeaway is that \emph{how} you prune matters far more than whether you prune at all. Random pruning actually performs worse than every standard MPPI configuration, including the smallest sample budget $K\!=\!50$---throwing out samples indiscriminately hurts the weighted average by dropping useful trajectories while keeping noisy ones.

Informed pruning, on the other hand, brings clear improvements. Top-Weight, Cost-Threshold, and Elite-CEM all reduce tracking error compared to unpruned MPPI $K\!=\!500$, though the gains vary across methods and tracks. All three achieve similar heading RMSE, which is expected since they are all selecting from the same pool of low-cost or high-weight samples. However, these heuristic methods still incur several boundary violations on both tracks, with the numbers getting worse on the lemniscate, where the narrowed self-intersection leaves less margin for error.

DM-fixed and DM-adapted consistently achieve the fewest boundary violations across both tracks. DM-adapted reaches the best or near-best tracking RMSE on both courses while keeping violations roughly an order of magnitude lower than the heuristic pruning baselines. The correlated noise, multi-iteration refinement, and influence-based pruning in DM-fixed produce smoother trajectories that naturally stay further from boundaries, while the adaptive mechanisms in DM-adapted further sharpen tracking by slowing down early in tight sections.

These trends hold across track geometries. The relative ordering of methods stays consistent on both tracks despite their different difficulties: the lemniscate's self-intersection and variable width demand tighter control, yet the DM-adapted method keeps violations near one on average, while heuristic pruning methods see their violation counts climb.

Finally, simply adding more samples shows diminishing returns—going from $K\!=\!200$ to $K\!=\!500$ barely changes tracking RMSE. Yet pruning $K\!=\!500$ down to 100 well-chosen samples beats the full unpruned budget. A smaller set of good samples outperforms a larger set of unfiltered ones.

\section{Conclusion}

We presented DM-MPPI, a framework that brings the datamodel idea from machine learning into MPPI control. We tested it on a kinematic bicycle model navigating two closed-loop tracks with obstacles. Informed sample pruning—keeping only the top $20\%$ of trajectories—substantially lowers tracking error compared to unpruned MPPI. Random pruning, by contrast, makes things worse than even the smallest sample budget. Top-weight, cost-threshold, and CEM-style pruning all improve tracking, but they still produce noticeably more boundary violations than the datamodel-based variants. DM-adapted consistently achieves the best or near-best tracking accuracy across both tracks while keeping boundary violations far below all other methods. This suggests that influence-aware selection offers a meaningful safety advantage over simple heuristic pruning.

\textbf{Future work} includes: extending evaluation to higher-dimensional systems and dynamic obstacles to test generality; replacing the linear datamodel with nonlinear or attention-based influence predictors that capture sample interactions; developing online adaptation to mitigate training distribution mismatch; integrating with formal safety mechanisms such as Control Barrier Functions~\cite{ames2017control} or safe importance sampling~\cite{gandhi2024safe} for hard constraint guaranties; coupling influence prediction with adaptive resampling, where the pruned budget is reallocated to resample in promising control regions; exploring how DM-MPPI can be combined with other MPPI improvements such as covariance adaptation and warm-starting.

\section*{Acknowledgment}
Claude was used to assist with the language editing of this manuscript.
 
\bibliographystyle{IEEEtran}
\bibliography{references}

\end{document}